# Self-Organizing Maps of Unbiased Ligand-Target Binding Pathways and Kinetics

Lara Callea[1], Camilla Caprai[2,3], Laura Bonati[1], Toni Giorgino[3*], Stefano Motta[1*]

[1] Department of Earth and Environmental Sciences, University of Milano-Bicocca, Piazza della Scienza 1, Milano, 20126, Italy.

[2] Department of Biosciences, University of Milan, via Celoria 26, Milan, 20133, Italy

[3] National Research Council of Italy, Biophysics Institute (CNR-IBF), Via Celoria 26, Milan, 20133, Italy

* Author to whom correspondence should be addressed: toni.giorgino@cnr.it, stefano.motta@unimib.it

## ABSTRACT

The interpretation of ligand-target interactions at atomistic resolution is central to most efforts in computational drug discovery and optimization. However, the highly dynamic nature of protein targets, as well as possible induced fit effects, makes difficult to sample many interactions effectively with docking studies or even with large-scale molecular dynamics (MD) simulations. We propose a novel application of Self-Organizing Maps (SOM) to address the sampling and dynamic mapping tasks, particularly in cases involving ligand flexibility and induced fit. The SOM approach offers a data-driven strategy to create a map of the interaction process and pathways based on unbiased MD. Furthermore, we show how the preliminary SOM mapping is complementary to kinetic analysis, both with the employment of network-based approaches and Markov State Models (MSM). We demonstrate the method by comprehensively mapping a large dataset of 640 µs of unbiased trajectories sampling the recognition process between the phosphorylated YEEI peptide and its high-specificity target Lck-SH2. The integration of SOM into unbiased simulation protocols significantly advances our understanding of the ligand binding mechanism. This approach serves as a potent tool for mapping intricate ligand-target interactions with unprecedented detail, thereby enhancing the characterization of kinetic properties crucial to drug design.

# INTRODUCTION

The elucidation of ligand-target interactions at the atomistic level is central to pharmaceutical research and drug development[1]. Despite the substantial progress achieved through conventional computational approaches such as docking studies, the challenges posed by the dynamic nature of protein targets and potential induced fit effects continue to impede reliable quantitative description of protein-ligand interactions, essential for drug discovery and optimization[2]. Molecular dynamics (MD) simulations, offering a dynamic and realistic depiction of biomolecular interactions by considering the inherent flexibility and adaptability of protein targets, have emerged as versatile tools to unravel the complexities associated with ligand binding[3,4]. Classical all-atom MD approaches, for example, model all of the atoms in a system as particles interacting via bonded and non-bonded potentials described by empirical force fields[5]. MD-based approaches overcome the limitations encountered in rigid docking models, allowing for a more accurate representation of induced fit effects and capturing subtle changes in protein conformation during ligand binding events. However, despite holding great promise, their effective utilization requires addressing challenges related to computational expense and the need for extensive sampling to comprehensively explore the conformational landscape[6,7]. This is particularly crucial for achieving atomistic resolution in ligand-target interactions, a prerequisite for robust drug discovery efforts.

Broadly speaking, MD-based methods can be divided into *biased* (or enhanced) and *unbiased* sampling approaches. Enhanced sampling methods address the issue of computational demands by applying perturbations to the Hamiltonian of the system enabling the reconstruction of free-energy landscapes and other properties[6,7]. Techniques such as alchemical free-energy perturbation[8], steered MD[9,10], metadynamics[11–13], Gaussian-accelerated MD[14], supervised MD[15], random acceleration MD[16], and many others[17–21] have enjoyed remarkable success, with the drawback that they usually rely on a choice of collective variables to be performed *a priori* and/or they introduce a bias that can influence the natural dynamics of the system. Recently, with the increase in computational power, long-time-scale processes have become computationally tractable with *unbiased* simulations[4,22,23]. In unbiased MD the system evolves only subject to force-field based potentials and thermostats, without any external bias, providing in principle a faithful representation of the system's dynamics. This characteristic enables the extraction of kinetic information, such as transition rates and timescales, directly from the trajectories. However, unbiased MD generates a large amount of data that is not straightforward to analyze. The calculation of kinetic properties in unbiased MD is facilitated by employing advanced analysis techniques, with Markov State Models (MSM) being a prominent example[24–26]. The construction of MSM, while a powerful tool for capturing the kinetic aspects of molecular processes, often encounters challenges. One notable difficulty arises from the sheer complexity and dimensionality of the data, making convergence a non-trivial task. The identification of suitable metrics for validation purposes further adds to the intricacy.

In response to these challenges, this paper proposes an application of Self-Organizing Maps (SOM)[27,28] as a novel strategy to account for the complexities associated with ligand flexibility and induced fit phenomena. We demonstrate the approach in a particularly challenging scenario, namely a flexible tetrapeptide ligand (pYEEI), in a high-specificity target (p56lck Src homology 2 SH2 domain). Src homology 2 (SH2) domains provide phosphorylation-dependent signaling receptor domains that recognize short peptides with very high sequence specificity and affinity[29]. The signaling function is of high physiological interest: at least one hundred human proteins contain an SH2 domain (InterPro), and phosphopeptide mimetics targeting SH2 have been explored to reduce proliferation in in-vitro breast cancer models, e.g. inhibiting the STAT3 pathway[30].

The SH2:pYEEI association occurs within a dynamic landscape on both interacting sides, making it very challenging to map out systematically. SOM offers unique clarity in understanding these complexities, providing a natural map of the interactive process and a robust means of verifying the outcomes. We demonstrate the method on a large dataset of 640 µs of unbiased trajectories, manyfold extending the ones used in previous studies[23]. Furthermore, we show how SOM mapping is highly suitable for kinetic analysis, facilitating an in-depth investigation of the temporal dynamics of the process. The method provides opportunities for several systematic kinetic analysis strategies, including both network-based communities and MSM. Of particular interest is the integration of SOM with MSM as an alternative to the initial clustering stage, thus offering the potential for a seamless integration of these two analytical paradigms. This integrative methodology enhances our understanding of ligand binding mechanisms and holds significant promise for optimizing the drug discovery process.

# METHODS

## Self-Organizing Maps

A Self-Organizing Map (SOM) is an unsupervised learning approach that facilitates the projection of high-dimensional data into a lower-dimensional space[27,31]. This technique has been widely used in biomolecular simulation analyses, with applications that includes clustering of conformations[32–35] for the analysis of pathways in enhanced sampling MD simulations[36–38]. For these purposes, we previously released PathDetect-SOM[39,40], a tool based on SOMs, that was here applied to a set of unbiased simulations of the SH2:pYEEI complex[23]. The SOM algorithm starts with the choice of a set of features that describes each data point (here a set of distances for each frame). Then the map is initialized and trained with the input vectors containing the values of the selected features for all the simulation frames. Each frame is considered as a data point and assigned to the neuron with most similar feature values. During the training process the feature values of a neuron and its neighbors are adjusted toward the values of the input vector assigned to that neuron. This process continues over multiple cycles to achieve an accurate low-dimensional representation of the data. The final prototype vector of each output neuron summarizes the conformations associated to the neuron and groups of similar conformations are mapped to neighboring neurons. In this work, the training was performed over 5000 cycles using a $20 \times 20$ sheet-shaped SOM with a hexagonal lattice shape and without periodicity across the boundaries. After training, each frame of the simulation set is assigned to a neuron on the map (hexagons), and each neuron represents a geometric microstate of the complex. In a second step, the neurons are further grouped in an optimal number of clusters by agglomerative hierarchical clustering, using Euclidean distances and complete linkage. The optimal number of clusters was selected according to the first maximum in the silhouette profile ranging in a reasonable interval (5-20 clusters). Properties such as the ligand RMSD from the X-ray structure[41] were displayed assigning to the neuron a color code proportional to the average value of the property for the frames belonging to that neuron.

## Network and connectivity analysis

An approximate transition matrix was estimated by counting the transitions between each pair of neurons in all the simulations. A graph was then built with nodes represented by neurons and edges were set to the negative logarithm of the transition probability between the corresponding neurons. For the sake of representation, transitions with fewer than 10 counts were not represented in the graph. For the analysis that make use of the transition probability matrix, the whole data was instead used. The distance between two nodes in the graph was calculated along the shortest path connecting them as the negative logarithm of the product of the pairwise transition probabilities between neurons along the path. A committor analysis was also performed computing the probability of hitting a set of states A before set B, starting from different initial states. In this case, the two extremes were the bound and unbound states, as detailed in the results section. To validate the obtained results, we employed a progressive bootstrap analysis to estimate the average standard deviation on the computed committor, as a function of the number of replicas used. We randomly selected subsets of replicates of increasing size. For each subset we performed 100 bootstrap resamples re-computing the transition matrix and estimating the average standard deviation of the committor computed on each neuron. Moreover, a bootstrap analysis was perfomed using 2/3 of the replicas, and the committor probability values were then obtained. This process was repeated 250 times, allowing the calculation of the average and the standard deviation on the obtained values. This was used to show the average value of the committor probability for each neuron and the standard deviation obtained from the bootstrap. All the analyses were performed in the R statistical environment using the kohonen[42,43], igraph[44] and markovchain[45] packages.

**Estimation of binding kinetics on-rates from trajectories**

The kinetics of the SH2:pYEEI association can be obtained from the distribution of times taken to reach the bound state. As a first approximation, one can model the process as a two-state irreversible transition between the unbound and the bound state; this model enables the estimation of the binding rate, assumed constant in time, from the ratio of binding events over the time sampled in the unbound state, after normalizing by the effective concentration[23]. Confidence intervals for the incidence rate (also known as Poisson rate exact confidence intervals) are then provided by Ulm's formula[46] (see Supplementary Methods).

MSM models of molecular systems rely on partitioning the conformational space into distinct states and estimate the transition probabilities between these states from simulated trajectories[47–49], Unbiased MD generates the necessary data for constructing a MSM, enabling a detailed understanding of the underlying kinetics of ligand-target interactions. The transition probability matrix can then be used to estimate thermodynamic (e.g. asymptotic state probabilities) and kinetic (e.g. rates) properties. The success of a Markovian description is sensitive to the precise partitioning of the state space: different partitions will be closer or further from markovianity[50], thus providing better or worse extrapolations on the long timescales. We therefore built a set of Markov models based on the states and communities identified by the SOM mapping procedure using the transition count estimators provided by the DeepTime Python library[51].

**The SH2-phosphopeptide system as a high-specificity recognition model**

The crystallographic structure of SH2 has been first reported by Waksman[52], who showed (albeit in a static structure) a complex two-pronged binding mechanism, likely occurring as a consequence of electrostatic steering at the N end of the peptide (namely pTyr), as well as induced fit in a hydrophobic region holding the C end. In a previous work, some of us[23] analyzed the dynamics of the two-pronged binding mechanism of the pYEEI peptide to the p56 lck SH2 domain through multiple parallel unbiased MD simulations, from which only five spontaneous binding events could be recovered[23]. The system has been modeled based on the crystal structure of human p56-lck tyrosine kinase SH2 in complex with the pYEEI phosphopeptide at 1 Å resolution by Tong et al. (PDB: 1LKK,[41]), where ligand was displaced by 40 Å to obtain an unbound starting conformation. The system was parametrized with the CHARMM27 force field. Water molecules present in the crystal structure were retained and the system was solvated in TIP3P explicit water and 150 mM NaCl, leaving a buffer of 52 Å of water around the protein in the direction of the ligand, and at least 12 Å in the other directions. Equilibration included 10 ns of simulation in the constant-pressure ensemble (Berendsen thermostat) followed by 20 ns in the constant-volume ensemble. The ligand was prevented from diffusing during equilibration by 1 kcal/mol/Å$^2$ harmonic restraints applied to its Cα atoms[53]. The resulting simulation box was $60 \times 66 \times 98$ Å³, with a smaller $40 \times 40 \times 60$ Å³ flat-bottom box restraining the center of mass of the ligand to a 20 mM effective concentration. The equilibrated systems were finally simulated with the ACEMD software on the GPUGRID.net distributed simulation network[53] in multiple replicas at 295 K, with particle-mesh Ewald treatment of long-range electrostatics. Further details of the simulation protocol are as previously reported[23]. In this work, to build a statistically-relevant structural decomposition of the process, we produced a dataset widely extending the previous analysis; the extended dataset built for this paper consists of 772 unbiased SH2:pYEEI MD trajectories, almost all 800 ns long (distribution in SI), with snapshots taken every 1 ns. The new dataset contains a total of 640 µs, extending the previous sampling over four-fold and enabling the use of statistical and graph-theoretical methods over the SOM map. The extended dataset is publicly available in full on Zenodo (see "Data Availability").

# RESULTS

## SOM Clustering of Unbiased Trajectories

The trained SOM is represented in Fig. 1, where each neuron (hexagon) corresponds to a configurational microstate, i.e. a peptide binding mode defined by specific values of the intermolecular distances used as input features. Neurons close to each other represent similar configurations. The map depicts a distribution of states ranging from the unbound state, (bottom left of the map) to the crystallographic-like bound state (top center) or alternative bound states (top right). This is due to the sampling of the binding process, providing sufficient representation to the macroscopically meaningful states, which are then captured during the training process. Hierarchical clustering grouped the 400 neurons into 11 clusters (represented with different colors on the map and labelled as A-K) that coarsely represent the binding geometries explored by the system during the simulations (Figure 1); one representative conformation from each cluster is shown to provide an overview of the corresponding macrostate. It is possible to identify a cluster representative of the unbound state, namely cluster A (blue in Figure 1); a series of clusters (B, C, D, F, G, H, J, K) describing the possible pre-bound states in which the peptide begins the first contacts with the protein; and two different bound states contained in clusters E and I (purple and yellow in Figure 1, respectively). Cluster E is characterized by conformations in which the ligand binds like in the crystallographic geometry, while cluster I represents an alternative bound state in which the ligand is rotated by 180° with respect to the X-ray geometry[41].

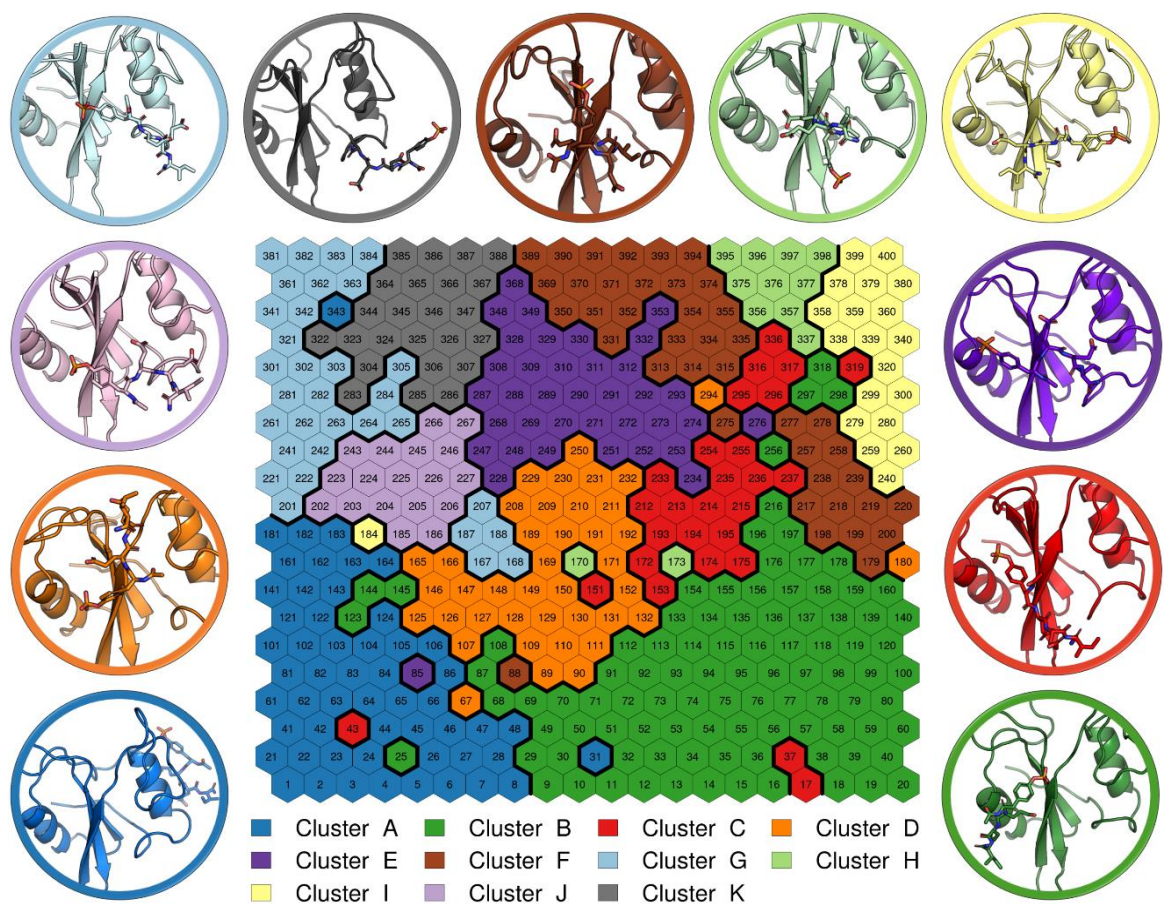

*Figure 1: SOM clustering of unbiased simulations of pYEEI binding to SH2 domain. The representative conformation of each cluster is depicted in cartoons with ligand in sticks.*

## Binding pathway analysis

Tracing the pathway followed by each trajectory on the SOM reveals a remarkable heterogeneity as each of them evolves following a different sequence of clusters. This is expected because, unlike simulations performed with MD methods in which a bias guides the system along a selected collective variable (CV), unbiased simulations evolve without following a specific direction, and in general explore transient pockets and kinetic traps[4,47].

All simulations start from the unbound state (cluster A), but only a minority reaches (or ends in) the crystallographic-like bound state (cluster E) within the sampled time. Plotting the average ligand RMSD values from the X-ray structure of the frames belonging to each neuron on the SOM (Figure 2), the lowest values are found for neurons 311, 291, 272 (purple circle in Figure 2), within cluster E, which we hence assume as the proper bound state. Twenty-two out of the 772 trajectories (2.8%) end in one of the three states. Using the three states as the definition of the bound state yields a $k_{on}$ of $3.5 \cdot 10^6$ s$^{-1}$ M$^{-1}$ (95% CI: 2.5 - 4.8 $\cdot 10^6$ s$^{-1}$ M$^{-1}$).

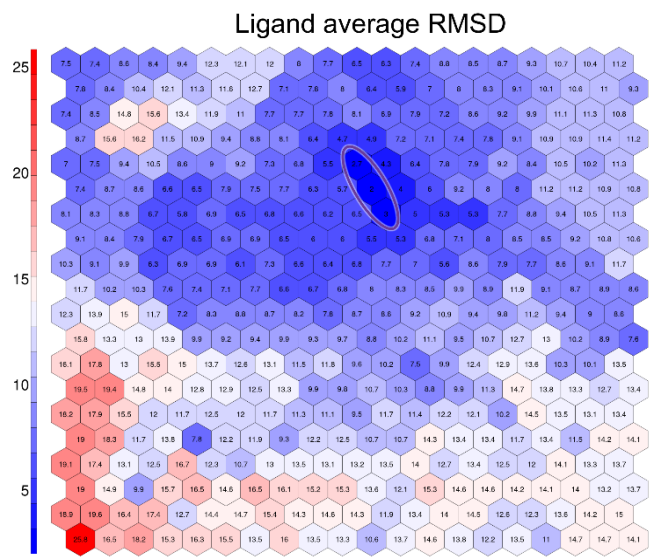

*Figure 2: SOM colored according to ligand average RMSD with respect to the bound state (values increasing from blue to red). States 272, 291, 311 (circled) have RMSD ≤ 3 Å.*

## The binding process as a transition network

The transition network analysis allowed us to summarize all the sampled pathways in a graph (Figure 3). All the simulations start from the unbound state (nodes in blue). Reading the transition network from the top, in yellow one finds nodes that lead to the alternative bound state in which the ligand is rotated 180° with respect to the X-ray geometry. Moving downwards from the unbound state toward the right side of the graph it is possible to observe the pathways that lead to the crystallographic-like bound state (purple nodes), and in particular toward the three neurons (311, 291 and 272) with the lowest ligand RMSD values from the X-ray structure.

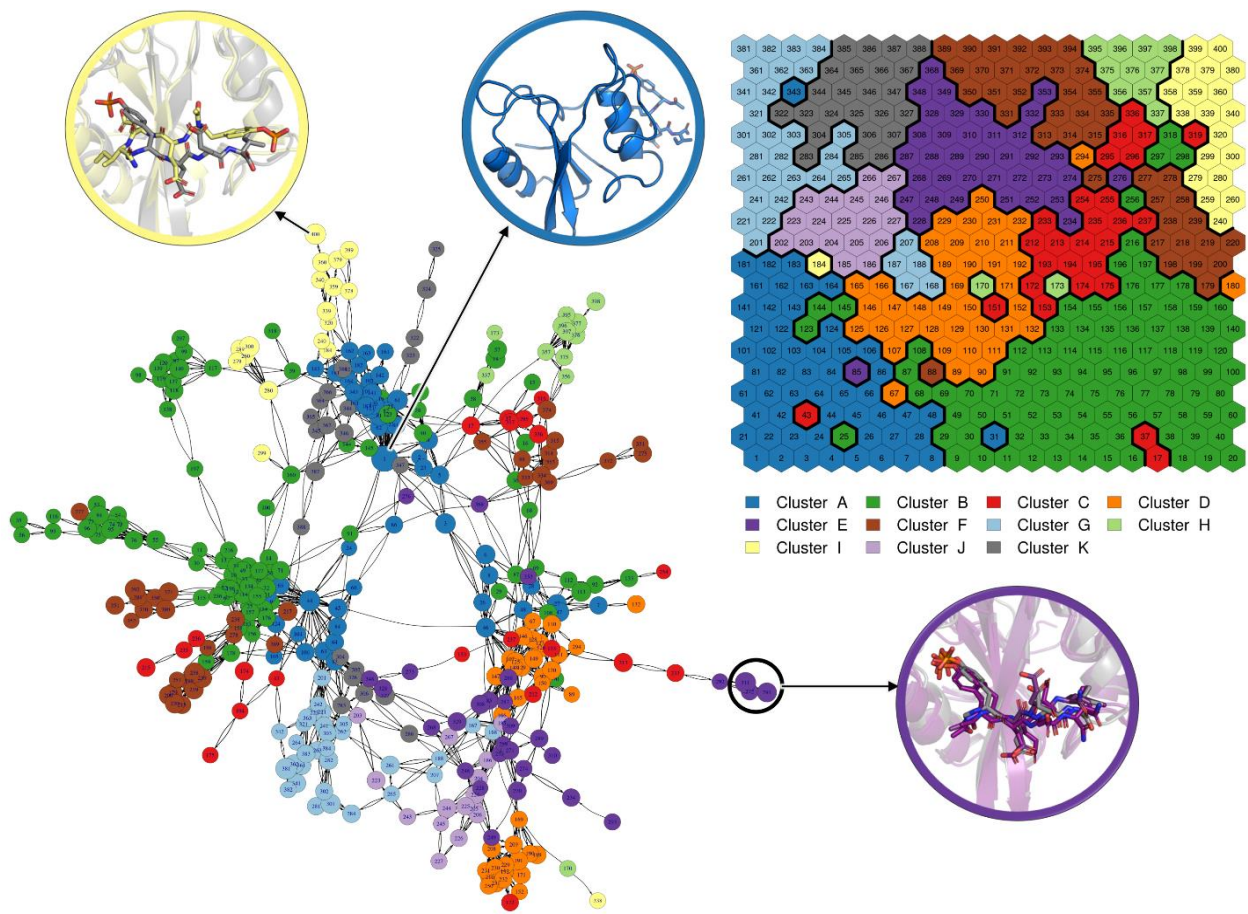

*Figure 3: Transition network analysis: transition network with nodes colored according to the SOM clusters. Representative conformations of neurons that characterize the unbound, crystallographic-like bound and alternative bound states are represented within circles using blue, magenta and yellow respectively for the cartoons and the ligand carbon atoms; cartoons and the ligand carbon atoms of the experimental structure, (PDB ID 1LKK[41]) is shown in gray.*

*Community detection identifies kinetically connected and transition states* - From the transition network, by analyzing the number of transitions between nodes, it is possible to detect "highly connected communities"[54]. Community analysis can be interpreted as a form of kinetic clustering, independent of any information about the bound and unbound states.

While geometric clustering yields information about conformational similarities by grouping together neurons having resembling features, community detection provides an overview of the kinetic relationships of the system. Community-based clustering was able to separate the nodes (neurons) belonging to the bound state from the surrounding ones, which the geometric clustering instead clumps together. This is evident by comparing the purple region (cluster E) in Figure 1, with the one highlighted in Figure 4, both corresponding to the crystallographic-like bound state in the two different clustering. Cluster E is significantly larger, including many more neurons than the community indicated in steel blue in Figure 4, which encompasses a group of only 6 nodes. Additionally, community analysis identifies the transition state (neuron 233), and places it in the same community as the bound state. Notably, a committor analysis (discussed in more detail later in text) confirms that neuron 233 is the closest to the iso-committor surface (committor value of 0.43), and therefore a likely representative of the transition state. The presence of a single node in the transition state may lead to the interpretation that this binding process can be indeed modeled as a two-state process with a downhill pathway after a single transition saddle point.

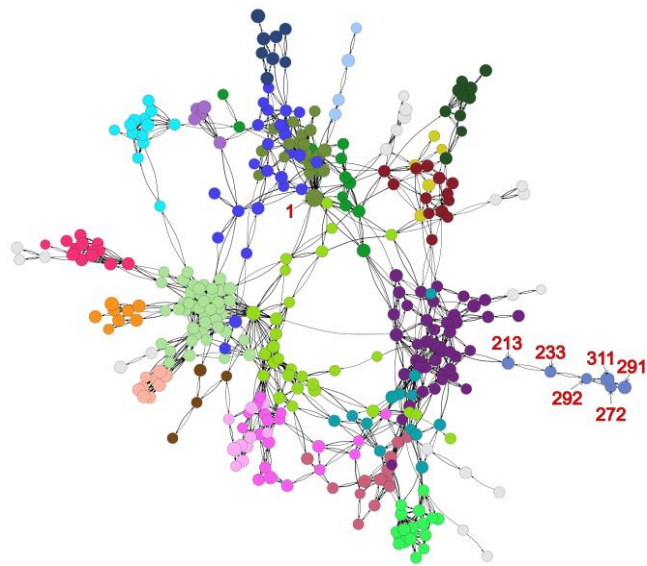

*Figure 4: Transition network colored according to the detected communities. Community detection identifies nodes (neurons) corresponding to the bound (272, 291, 311) and transition (233) states, highlighted in steel blue. Nodes with insufficient transitions to be correctly assigned to a community are grayed out.*

*Shortest-path analysis shows the prototypical binding pathway* **-** To better understand the steps involved in the pYEEI peptide binding to the SH2 domain, we computed the sub-optimal (yellow in Figure 5) and the shortest (orange in Figure 5) pathways linking the unbound state (neuron 1) with the crystallographic-like bound state (neuron 291). The shortest path is defined as the one that minimizes the cumulative sum of edge weights between the starting node and the ending node.

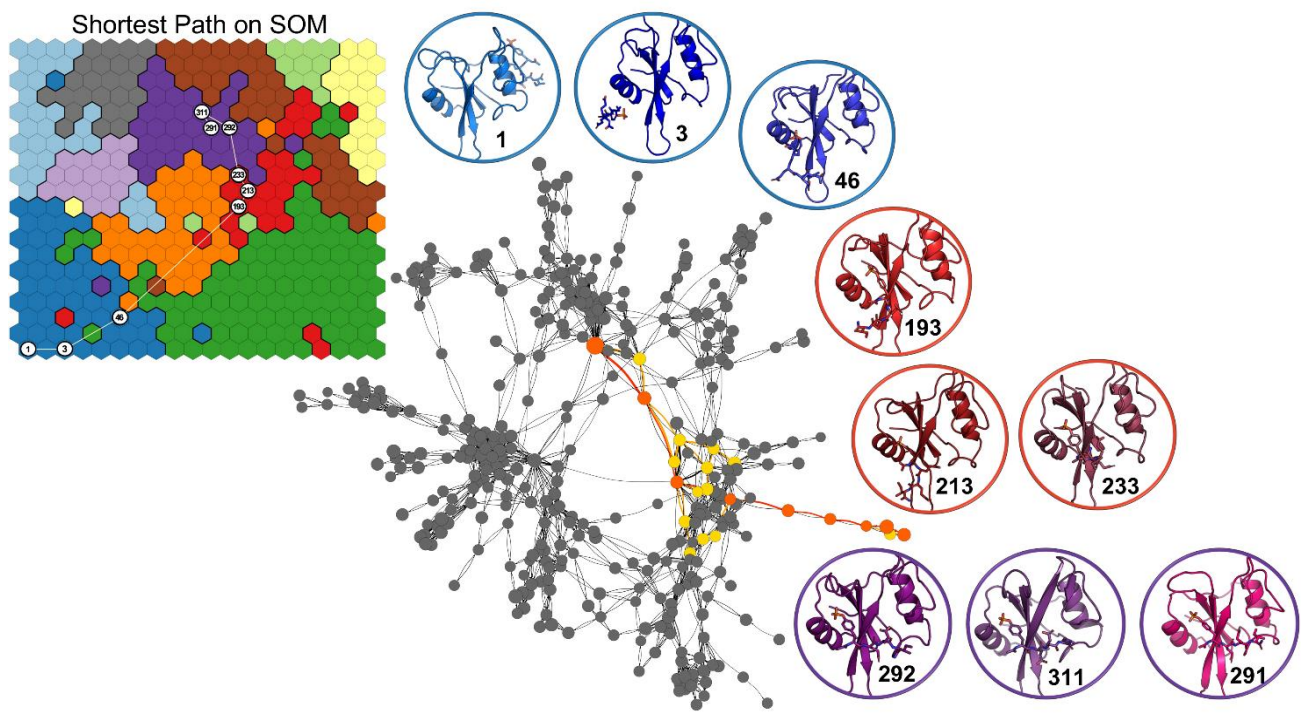

*Figure 5: Pathways analysis on the SOM and in the transition network. All sub-optimal pathways are plotted in the transition network with nodes colored in yellow. The shortest path is mapped on the graph with orange nodes and marked on the SOM with white circles. The representative conformations of neurons that characterize the shortest path are depicted in cartoons with the ligand in sticks.*

Furthermore, we compared the previously proposed mechanism[23] for peptide binding with the present proposal based on the analysis of the shortest path. As shown in Figure 6, initially the peptide is in the bulk (a). The first step is characterized by the initial contact of the pY group of the peptide with the protein, which involves the formation of two salt bridges with residues R154 and K182 (b). This interaction is further stabilized in the subsequent step by the formation of a hydrogen-bond network with residues S156, S158, and A160, that traps pY in the bound state (c). The following step involves the contact of the second E residue through a salt bridge with R184 (d). Next, the first E residue of the peptide stabilizes in the bound position, forming a hydrophobic interaction with both K179 and Y181 (e). The final step for peptide binding involves positioning of the I residue between the EF and BG loops, favored by a network of hydrophobic interactions with residues I193, Y209, and L216 (f). Based on the above discussion, the two mechanisms appear to be consistent.

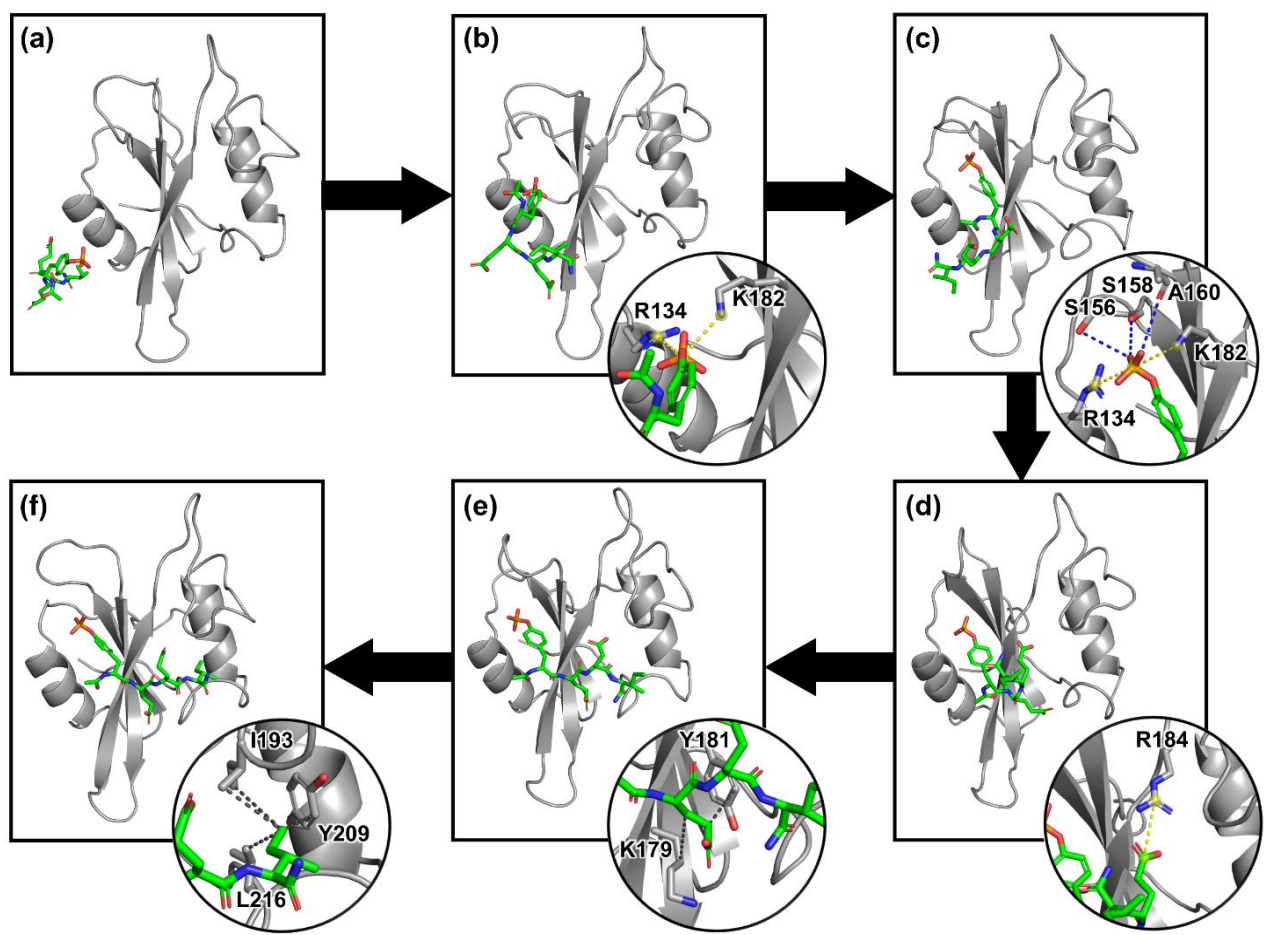

*Figure 6: Insight into the peptide binding mechanism: schematic representation of the different steps Ranging from unbound (a); initial contact (b); anchoring of pTyr (c); salt bridge of the second Glu residue with R184 (d); positioning of the second Glu residue (e); positioning of the C-terminal residue of the peptide between the EF and BG loops (f). In cartoon and gray sticks the protein and the residues involved in the interaction with the pYEEI peptide, represented in green stick. In yellow dashes salt bridges, in blue dashes hydrogen bonds, in dark gray dashes the hydrophobic interactions.*

*Bound-state heterogeneity* - All simulations reaching the bound state showed that the bound geometry is characterized by different conformations well represented by neurons 291 and 311, which exhibit similar and very low ligand RMSD values to the X-ray structure. Interactions characteristics of the two microstates are slightly different: neuron 311 shows a salt bridge with K179, that is lacking in neuron 291, and neuron 291 presents a more extended network of hydrophobic interactions involving the EF and BG loops (as shown in Figure S2). The fact that different simulations reach bound states with slightly different conformations is expected, as individual replicas may explore distinct, yet closely related, conformational states that however share the same key interactions. Upon plotting the number of simulations that ended in each neuron on the SOM (left-hand panel of Figure 7), we observed that, after neuron 1 (representing the unbound state), neuron 311 exhibited the highest occurrence, capturing the largest number of replicas. This analysis also revealed that neurons 381 and 400 (at the top left and right vertices of the SOM), neuron 351 (just above neuron 311), and neuron 55 (at the bottom right corner) have from 8 to 10 simulations ending in them, suggesting that they could be kinetic traps. Section "Markovian Modeling of the Binding Process" will provide estimates of the asymptotic (equilibrium) probabilities of each state computed through Markov state models.

## Committor analysis identifies the transition state

We conducted a committor analysis (right-hand panel of Figure 7) to calculate the probability of the system to access the bound conformation (neuron 311) before reaching the unbound conformation (neuron 1), starting from each neuron. The convergence of the calculation was assessed by evaluating the average standard deviation of the computed committors through a progressive bootstrap analysis (see Methods section for further details and Figure S3). The analysis revealed that once the number of replicas reaches approximately 500, the average standard deviation of the committor probabilities stabilizes, confirming the convergence of the calculation. As the transition matrix was constructed from unbiased simulations, conformations with a committor of around 0.50 can be considered in proximity of the transition states. In this case, the energy barrier appears to be located around neuron 233 (committor value 0.43), in the conformations of which the pY group of the peptide is fixed to the protein and a salt bridge is formed between the second E residue of the peptide and R184.

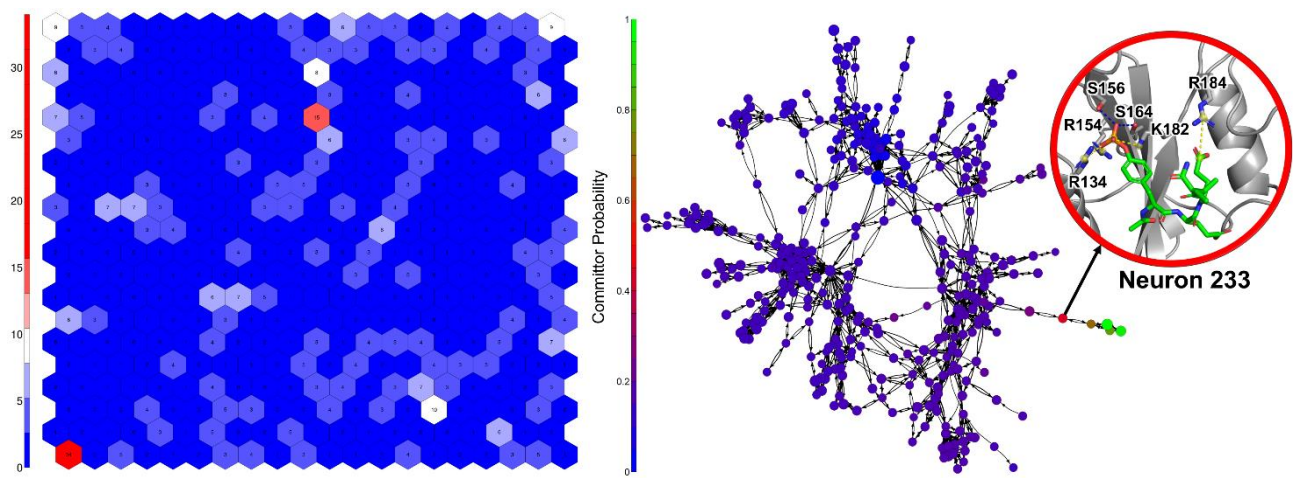

*Figure 7: On the left, SOM is annotated with the number of simulations ending in each neuron (values increasing from blue to red); on the right, graph network with nodes colored according to the committor*

*probability analysis (increasing values from blue to green). The representative conformation of neuron 233 is depicted in cartoons with the ligand and the residues involved in the interactions in green and gray sticks, respectively. To verify the robustness of our model and validate the previously discussed results, we conducted a bootstrap analysis (see Methods section and Figure S4).*

The barrier appears to be due to the probability that the collision between the two molecules occurs with the correct orientation, allowing the first salt-bridge to form properly and aligning the rest of the molecule correctly to form the second salt-bridge. For the pYEEI peptide binding to the SH2 domain, the initial encounter must position the tyrosine-phosphate (pY) group to form the first salt bridge with residues R154 and K182 on the protein surface. This initial contact is crucial, as it stabilizes the initial complex and orients the peptide for subsequent binding steps. If the peptide collides in an incorrect orientation, the necessary interactions will not form, and the peptide may dissociate or bind in a less favorable mode. This requirement for precise orientation introduces an entropic cost. The system must overcome the disorder associated with the numerous possible orientations during the peptide's diffusion. The correct alignment is less probable, contributing to the observed transition state at neuron 233, identified by committor analysis. This state marks a critical point where the probability of progressing to the bound state equals that of returning to the unbound state, indicating a significant entropic barrier.

## Communities provide a robust basis to compute the binding kinetics

The computation of the binding kinetics based on atomistic data is sensitive to the precise definition of what microstates constitute "the bound state". In this respect, the community analysis graph provides a very robust approach for the computation of kinetics. First, we assume a threshold of $\varepsilon = 0.5$ for the committor value $c$, and we consider all the states whose committor is $c_i \geq \varepsilon$ as the bound state (otherwise being unbound). This yields 52 binding events and therefore a $k_{on}$ rate of 4.4 (95% CI 3.3 – 5.8) · $10^6$ $s^{-1}$ $M^{-1}$. Taking the transition state at $c = 0.5$ is a common assumption; however, it is interesting to do a sensitivity analysis to study how strongly the $k_{on}$ depends on the precise value of the threshold. We again computed $k_{on}$ as the event rate, i.e. the ratio between the number of binding events observed to the total unbound sampled time, now as a function of $\varepsilon$ (Figure 8). The rates are insensitive to the precise choice of $\varepsilon$, being essentially constant as for the canonical $\varepsilon = 0.5$ in a wide range of values, thus again confirming that the community analysis projects the complex process into a good reaction coordinate (albeit discretized) with two attraction basins separated by a well-defined transition region.

This contrasts, for example, with an RMSD-based definition computing the similarity of the configuration of the ligand in the final frame of each simulation to the (assumed known) X-ray structure. Such a definition would imply, e.g., that 24 simulations (≈3%) end in a frame with an RMSD ≤ 2 Å, while 143 simulations (≈18%) end in a state with an RMSD ≤ 5 Å. By taking an RMSD threshold between 1.8 and 3.0 Å as a definition of the bound state, the sensitivity analysis of $k_{on}$ identifies binding rates that are relatively constant (Supplementary Figure S5). Using the larger plateau region between 1.8 and 2.4 Å, one obtains a $k_{on}$ rate of 2.0 (95% CI: 1.3-3.0) · $10^6$ $s^{-1}$ $M^{-1}$. The RMSD-based approach to the definition of the bound state however has the drawback that the $k_{on}$ value depends strongly both on (1) the precise knowledge of the bound structure, as well as (2) the RMSD threshold chosen. A Other geometric observables could, in principle, be similarly used to define the kinetics; e.g. the fraction of native contacts provides a similar value for the $k_{on}$ – albeit with a plateau too narrow to identify reliably (Supplementary Figure S6). To assess the impact of the simulation size on the estimated $k_{on}$, we calculated $k_{on}$ as a function of replica length (Figure S7, top), and the number of replicas (Figure S7, bottom). The results show that after approximately 350 ns per replica (total aggregate simulation time of 280 μs), the estimated $k_{on}$ reaches a plateau, indicating the convergence of the calculation.

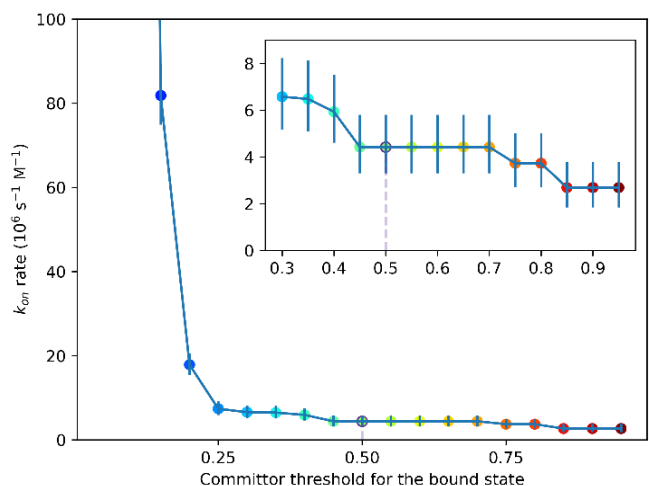

*Figure 8: Sensitivity analysis of the on-rate of the SH2:pYEEI binding process as a function of the committor value used to define the bound state. The community-based projection makes the $k_{on}$ rate remarkably robust to the definition of the bound state (committor threshold, on the horizontal axis; the iso-committor value of 0.5 is highlighted).*

## Markovian Modeling of the Binding Process

The SOM-based discretization lends itself to be a basis for a Markovian modeling of the association process. An MSM analysis can be conducted either discretizing the state space according to the SOM neurons, or according to the communities previously identified by community analysis; in both cases the implied timescales are quite similar (Figure S8), pointing to relaxation processes on the order of 10 μs and 2 μs respectively, around a lag time of 300 ns. MSM defined based on communities displays convergence at slightly lower lag times, despite the coarser states.

We used the MSMs to extrapolate the asymptotic (equilibrium) state probabilities. Equilibrium probabilities have a distinct peak in the 272, 291, 311 neuron triplet, clearly identifying the bound state. The three states collectively account for 54% of the equilibrium probability distribution. Boltzmann inversion of the equilibrium probabilities provides free energy values. Minor free energy minima are at neurons 400 (reverse-bound, cluster I), neuron 381 ("vertically-bound", cluster G), neuron 392 ("L-shaped", cluster F) and few other likely metastable configurations previously identified and discussed in the SOM.

Finally, even though this is not the main objective of the analysis, equilibrium probabilities can be converted into standard free energies of binding by Boltzmann inversion and accounting for the ligand concentration *versus* the standard state. Figure 9 shows a free energy landscape of the binding process mapped on the SOM neurons, reconstructed by building a MSM on the whole set of unbiased trajectories, which clearly identifies the bound triplet.

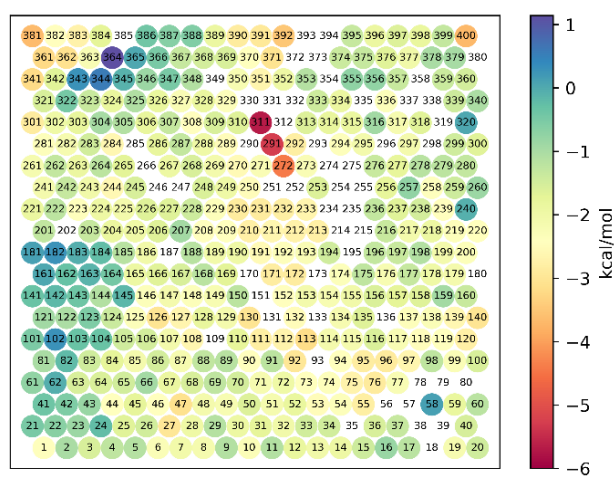

*Figure 9: Standard free energy of each SOM neuron, obtained by Boltzmann-inversion of the equilibrium probabilities estimated Markov-modeling the binding process at a lag time of 350 ns. Neurons of the bound*

*state (272, 291, 311) have the lowest free energy values with a total equilibrium probability of 0.54; neurons not belonging to any community are left blank. The zero of the free energy is assigned to the state unbound at 1 M; it was set applying the concentration correction (-2.4 kcal/mol) to state 1, namely unbound at simulation concentration.*

## CONCLUSIONS

In this work, for the first time, we applied SOM to an extensive dataset of SH2-pYEEI binding trajectories obtained from unbiased MD simulations. This approach provides valuable insights into the evolution of the system by revealing geometric clusters. Modeling the binding process as a transition network enabled the identification of key kinetic stages like the transition state, the bound state, and potential kinetic traps. A simple two-state treatment of the reaction process projected on the SOM discretization yields association rates in agreement with the experimental values. The combination of Markov models with SOM discretization yielded a robust free energy landscape, clearly pinpointing the bound state as the most energetically favorable configuration. This combined framework provides an accurate description of complex processes characterized by heterogeneous pathways. The results of this method are especially notable as they account for the remarkable flexibility of the tetrapeptide and the proper accounting of induced fit effects on the receptor.

A key advantage of using SOM is its ability to preserve the topological relationships between microstates, which enhances the interpretability of the conformational landscape compared to other traditional clustering methods. Other dimensionality reduction methods however exist such as Principal Component Analysis (PCA). PCA reduces dimensionality by focusing on the variance captured by the first few principal components, but it can overlook subtle yet important features of the data. In contrast, SOMs create a map where, at the same time, data are grouped in microstates and similar data are spatially close, providing a more intuitive and detailed representation. This allows for the detection of geometric patterns and the subsequent construction of a transition network that accurately captures the kinetics of the binding process. The SOM-based approach not only facilitates the identification of key states but also offers a versatile platform for further analysis, such as mapping external properties or performing kinetic clustering, which PCA alone does not inherently provide. Other approaches for pathway detection in MD simulations exist such as the one presented in[55]. However, these methods are not intended to cluster frames and obtain microstates that forms the unique platform offered by SOM for the subsequent analysis.

Despite the significant advantages of SOM, it is not without limitations. In this particularly challenging case, one major restriction is that – due to the inherent flexibility of both the protein and peptide, combined with the limited number of simulations reaching the bound state – we employed a set of "non-blind" distances to enhance the description of the correct bound state. While SOM can operate in a "blind" mode using all intermolecular distances, this approach tends to treat all sampled states equally. However, for this complex scenario, a more selective approach was necessary. Additionally, a careful interpretation of the free energy values obtained from the MSM analysis is needed for two key reasons. First, each neuron in the model represents a subset of the phase space of different volume. Second, the equilibrium probabilities identified by MSM are likely underestimated due to insufficient sampling near the bound state, which can be addressed through adaptive sampling schemes.

An additional limitation of the approach presented is the extensive sampling required to obtain reliable statistics on the binding process. While unbiased MD simulations are straightforward, requiring no collective variable definitions and avoiding external biases that could alter the physics of the process, they demand a substantial amount of simulation time, which may be prohibitive for large-scale drug design campaigns. However, the continuous optimization of MD software and the increasing computational power have led to orders-of-magnitude increases in the feasible simulation lengths over the past decade. This trend is expected to continue with advancements in artificial intelligence and quantum computing, making studies like the one presented here more routine in the future. In the context of a drug design campaign, it remains challenging to study a large number of ligands using this approach due to its high computational cost. Nevertheless, analyzing one or a few highly promising ligands could be valuable in identifying the key characteristics that make a ligand attractive or in uncovering potential barriers during the binding process that could be mitigated through proper functionalization, ultimately enhancing the ligand's efficacy.

Finally, while the data-driven approach we presented here has promising implications in drug design and optimization tasks, for pharmaceutical purposes one may want to investigate the off-rate, which has an important correlation to biological activity[56–58]. Sampling $k_{off}$ (or residence times) with purely unbiased approaches like the one presented in this paper is much more challenging than $k_{on}$ because of the far longer timescales involved. While feasible for shorter residence times (hundreds of μs are becoming accessible)[59,60] biased sampling approaches like (infrequent) meta dynamics[61], adaptive sampling markov models[62], weighted ensemble[63] and similar may still be approaches of preference for this task. Future work will focus on characterizing and reducing the sensitivity of the method to the a priori knowledge of the bound pose, thereby broadening its applicability to even more challenging "blind pathway reconstruction" tasks able to simultaneously recover pose, binding paths, and kinetics. This could be achieved e.g. though the integration of SOM with an adaptive seeding technique to enrich the number of transitions between neurons to improve the robustness and accuracy of the analysis.

## SUPPLEMENTARY MATERIAL

The supplementary material contains additional information and analyses supporting the main text. It includes figures detailing the selected atoms for interatomic distance calculation and used for SOM training (Figure S1), the three-dimensional structures of representative conformations for neurons 291 and 311 with annotated interactions (Figure S2), value of the average standard deviation of the computed committors in a progressive bootstrap analysis (Figure S3) and the average values and standard deviation for committor probabilities obtained from bootstrap analysis (Figure S4). Sensitivity analyses of the on-rate of the SH2 binding process are provided, examining the influence of the RMSD threshold value (Figure S5) and the fraction of crystallographic native contacts (Figure S6). Analysis of the *on* rate of the SH2:pYEEI binding process as a function of the simulation time and the number of replicas (Figure S7). Implied timescales for a Markov State Model built on SOM neurons and communities, showing convergence and relaxation processes, are also included (Figure S8). Supplementary methods detail the Poisson rate and confidence interval calculations, providing the formula for estimating incidence rates in a two-state irreversible model, along with the corresponding confidence intervals using Ulm's formula (Poisson rate confidence interval).

## ACKNOWLEDGMENTS

TG acknowledges funding from the Spoke 7 of the National Centre for HPC, Big Data and Quantum Computing (CN00000013) and from the PRIN 2022 (BioCat4BioPol) from the Ministero dell'Università e Ricerca, funded by the European Union – NextGenerationEU. TG thanks the volunteers of GPUGRID.net for donating computing time for the simulations. We acknowledge CINECA awards under the ISCRA initiative and the agreement with the University of Milano-Bicocca, for the availability of high-performance computing resources and support.

# AUTHOR DECLARATIONS

## Conflict of Interest

The authors have no conflicts to disclose.

## Author Contributions

**Lara Callea:** Data curation, Formal Analysis, Investigation, Visualization, Writing– original draft, Writing– review & editing; **Camilla Caprai:** Visualization, Writing– review & editing. **Laura Bonati:** Conceptualization, Supervision, Writing– review & editing; **Toni Giorgino:** Conceptualization, Supervision, Formal analysis, Writing/Review & Editing; **Stefano Motta:** Conceptualization, Supervision, Methodology, Writing/Review & Editing.

# DATA AVAILABILITY

The full dataset supporting the findings of this study are openly available in Zenodo at http://doi.org/10.5281/zenodo.12205888.